\documentstyle[12pt]{article}

\begin{document}


\def\a{\alpha}
\def\b{\beta}
\def\c{\varepsilon}
\def\d{\delta}
\def\e{\epsilon}
\def\f{\phi}
\def\g{\gamma}
\def\h{\theta}
\def\k{\kappa}
\def\l{\lambda}
\def\m{\mu}
\def\n{\nu}
\def\p{\psi}
\def\q{\partial}
\def\r{\rho}
\def\s{\sigma}
\def\t{\tau}
\def\u{\upsilon}
\def\v{\varphi}
\def\w{\omega}
\def\x{\xi}
\def\y{\eta}
\def\z{\zeta}
\def\D{\Delta}
\def\G{\Gamma}
\def\L{\Lambda}
\def\F{\Phi}
\def\P{\Psi}
\def\S{\Sigma}

\def\o{\over}

\def\IJMP{Int.~J.~Mod.~Phys. }
\def\MPL{Mod.~Phys.~Lett. }
\def\NP{Nucl.~Phys. }
\def\PL{Phys.~Lett. }
\def\PR{Phys.~Rev. }
\def\PRL{Phys.~Rev.~Lett. }
\def\PTP{Prog.~Theor.~Phys. }
\def\ZP{Z.~Phys. }

\def\beq{\begin{equation}}
\def\eeq{\end{equation}}


\title{
  \begin{flushright}
    \large UT-786
  \end{flushright}
  \vspace{5ex}
  Supersymmetry-Breaking Models \\ of Inflation}
\author{Izawa K.-I. \\
  \\  Department of Physics, University of Tokyo \\
  Tokyo 113, Japan}
\date{August, 1997}
\maketitle

\begin{abstract}
We consider dynamical models of supersymmetry breaking
which naturally incorporate cosmological inflation.
The inflaton plays an inevitable role in dynamical SUSY breaking
and the hierarchical scales of inflation and SUSY breaking
are simultaneously realized through a single dynamics.
\end{abstract}

\newpage

Supersymmetry, if present in nature, is broken at low energy
so that the observed particles do not
have superpartners with the same masses.
Spontaneous SUSY breaking at low energy may
be realized through dynamics of nonabelian gauge theory.
The vector-like model of dynamical SUSY breaking
\cite{Iza}
provides one of the simplest models that serve as realistic
SUSY-breaking sectors
\cite{Yan}
of the SUSY standard model.

In this paper, we provide a slight modification of the vector-like model
of dynamical SUSY breaking,
which naturally incorporates cosmological inflation of the hybrid type
\cite{Lyt}.
The inflaton plays an inevitable role in dynamical SUSY breaking
and the hierarchical scales of inflation and SUSY breaking
are simultaneously realized through a single dynamics.

Let us consider a SUSY SU(2) gauge theory with four doublet chiral
superfields $Q_i$ and six singlet ones $Y^a$ and $Z$.
Here $i$ and $a$ denote the flavor indices
($i = 1, \cdots, 4; \  a = 1, \cdots, 5$).
Without a superpotential, this model has a flavor SU(4)$_F$ symmetry.
We impose a flavor SP(4)$_F$ symmetry on our model,
where we adopt a notation ${\rm SP}(4)_F \subset {\rm SU}(4)_F$.
The superpotential of our model is given by
\begin{equation}
  W = W_Y + W_Z;
  \quad W_Y = \l_Y Y^a (QQ)_a, \  W_Z = {\l_Z \o M^{2(n-1)}}Z(QQ)^n,
\end{equation}
where $\l_Y$ and $\l_Z$ denote coupling constants of order one
and $(QQ)_a$ and $(QQ)$ denote a five-dimensional representation
and a singlet of SP(4)$_F$, respectively, in the gauge invariants
$Q_i Q_j$, which constitute a six-dimensional representation of SU(4)$_F$.
We take the reduced Planck scale $M$ as a natural cutoff in supergravity,
which is assumed in our construction.

SUSY breaking is realized in the present model
with the same dynamics investigated in Ref.\cite{Iza}.
Indeed, for $n=1$, the model is none other than the vector-like model
of dynamical SUSY breaking.
We henceforth put $n > 1$ to achieve a hierarchical structure
of the physical scales in the model.%
\footnote{If one would like to avoid a nonrenormalizable interaction
in the superpotential $W$,
one may consider the case $n = 1$ with a tiny coupling $|\l_Z| \ll 1$.}
The SUSY-breaking scale is given by $F_Z \simeq \l_Z \L^{2n} / M^{2(n-1)}$,
where $\L$ is a dynamical scale of the SU(2) gauge interaction.
Note that the superpotential $W_Y$ plays a crucial role
to cause dynamical SUSY breaking in the present model.

In a regime $|Y^a| \gg \L$, the effective superpotential of the model
is given by
\cite{Aff}
\beq
  W_{eff} \simeq \l_Y \L^2 \sqrt{Y^a Y^a},
\eeq
and the effective K\"ahler potential is approximately canonical.
Hence the effective potential is almost flat along $Y^a$
with $V_{eff} \simeq |\l_Y \L^2|^2$.
This is appropriate for inflationary dynamics.%
\footnote{A similar dynamics was considered in Ref.\cite{Dim}
to generate a flat potential for inflation.}
In fact, the radiative and supergravity corrections to the effective potential
allow the field space $|Y^a| \leq M$ to possess a region large enough
which satisfies the slow-roll condition for inflation
\cite{Lyt, Dim}.
The point is that the scale of inflation can be much larger than that of
the SUSY breaking in our model, which is suitable for
realistic primordial inflation.
It is remarkable that the model for SUSY breaking automatically
contains an inflationary sector $W_Y$ in the present construction.

In the above model, the singlet $Z$ contains a light scalar component
compared with the SUSY-breaking scale.
We may make it heavier through radiative corrections
\cite{Din}
by means of a superpotential
\beq
  W'_Z = Z[{\l_Z \o M^{2(n-1)}}(QQ)^n - \l_X X^2] + {f \o M^{2n-1}}(QQ)^mZ'X
\eeq
instead of $W_Z$ in the superpotential $W$,
where we have introduced additional singlets $X$ and $Z'$.

We finally consider a variant of the above model which incorporates
`new inflation' rather than SUSY breaking.
Let us take a superpotential
\beq
  W''_Z = {\l_Z \o M^{2(n-1)}}(QQ)^n Z - {f \o M^2} Z^5
\eeq
instead of $W_Z$ in the superpotential $W$.
Then the sector given by $W_Y$ may yield `pre-inflation'
\cite{Kaw},
which dynamically realizes the initial condition for `new inflation'
\cite{New}
of the inflaton $Z$.%
\footnote{We considered the case $n = 2$ as a dynamical model
of `new inflation' in Ref.\cite{Kaw}, which is to be supplemented by
a separate sector of `pre-inflation'.
We here claim that the dynamical model for `new inflation'
automatically contains a `pre-inflation' sector in itself,
and thus it requires no additional sector of `pre-inflation'.}
The point here is again that the scale of `pre-inflation'
can be much larger than that of `new inflation' in the model,
which leads to the initial condition for realistic `new inflation'.

We have presented a few models with two hierarchical scales
of physical interest generated naturally by a single dynamics.
Along similar lines of construction, we may consider various models
with two or more physical scales with large hierarchy.

\section*{Acknowledgement}

The author would like to thank T.~Yanagida for valuable discussions.


\begin{thebibliography}{99}

\bibitem{Iza}
  Izawa~K.-I. and T.~Yanagida, \PTP {\bf 95} (1996) 829; \\
  K.~Intriligator and S.~Thomas, \NP {\bf B473} (1996) 121.

\bibitem{Yan}
  Izawa~K.-I. and T.~Yanagida, \PTP {\bf 94} (1995) 1105; \\
  T.~Hotta, Izawa~K.-I., and T.~Yanagida, \PR {\bf D55} (1996) 415; \\
  Izawa~K.-I., hep-ph/9704382; \\
  Izawa~K.-I., Y.~Nomura, K.~Tobe, and T.~Yanagida, hep-ph/9705228; \\
  See also H.~Murayama, hep-ph/9705271; \\
  S.~Dimopoulos, G.~Dvali, R.~Rattazzi, and G.F.~Giudice, hep-ph/9705307; \\
  M.A.~Luty, hep-ph/9706554; \\
  S.~Dimopoulos, G.~Dvali, and R.~Rattazzi, hep-ph/9707537.

\bibitem{Lyt}
  For a review, D.H.~Lyth, hep-ph/9609431.

\bibitem{Aff}
  I.~Affleck, M.~Dine, and N.~Seiberg, \NP {\bf B256} (1985) 557; \\
  N.~Seiberg, \PL {\bf B318} (1993) 469.

\bibitem{Dim}
  S.~Dimopoulos, G.~Dvali, and R.~Rattazzi, hep-ph/9705348.

\bibitem{Din}
  M.~Dine, W.~Fischler, and D.~Nemeschansky, \PL {\bf B136} (1984) 169.

\bibitem{Kaw}
  Izawa~K.-I., M.~Kawasaki, and T.~Yanagida, hep-ph/9707201.

\bibitem{New}
  Izawa~K.-I. and T.~Yanagida, \PL {\bf B393} (1997) 331.

\end{thebibliography}
\end{document}